\documentclass{sigchi}

\toappear{ACM acknowledges that this contribution was authored or co-authored by an employee, contractor or affiliate of the national government. As such, the Government retains a nonexclusive, royalty-free right to publish or reproduce this article, or to allow others to do so, for Government purposes only.\\
{\emph{UIST'14}}, October 05--08 2014, Honolulu, HI, USA. \\
Copyright is held by the owner/author(s). Publication rights licensed to ACM. \\
ACM 978-1-4503-3069-5/14/10\ ...\$15.00.\\
http://dx.doi.org/10.1145/2642918.2647368}

\def\plaintitle{Teegi: Tangible EEG Interface}
\def\plainauthor{Jérémy Frey, Renaud Gervais, Stéphanie Fleck, Fabien Lotte and Martin Hachet}
\def\plainkeywords{Tangible Interaction; EEG; Spatial Augmented Reality; Learning}

\usepackage{balance}  
\usepackage{graphics} 
\usepackage{times}    
\usepackage{url}      

\makeatletter
\def\url@leostyle{%
  \@ifundefined{selectfont}{\def\UrlFont{\sf}}{\def\UrlFont{\small\bf\ttfamily}}}
\makeatother
\def\pprw{8.5in}
\def\pprh{11in}

\usepackage[pdftex]{hyperref}
\hypersetup{
pdftitle={\plaintitle},
pdfauthor={\plainauthor},  
pdfkeywords={\plainkeywords},
bookmarksnumbered,
pdfstartview={FitH},
colorlinks,
citecolor=black,
filecolor=black,
linkcolor=black,
urlcolor=black,
breaklinks=true,
}

\usepackage[utf8]{inputenc}

\begin{document}

\title{\plaintitle}

\numberofauthors{5}
\author{
  \alignauthor Jérémy Frey$^{*}$\\
    \affaddr{Univ. Bordeaux}\\
    \email{jeremy.frey@inria.fr}\\
  \alignauthor Renaud Gervais$^{*}$\\
    \affaddr{Inria}\\
    \email{renaud.gervais@inria.fr}\\ 
  \alignauthor Stéphanie Fleck\\
    \affaddr{Univ. Lorraine}\\
    \email{stephanie.fleck@univ-lorraine.fr}\\
  \alignauthor Fabien Lotte\\
    \affaddr{Inria}\\
    \email{fabien.lotte@inria.fr}\\
  \alignauthor Martin Hachet\\
    \affaddr{Inria}\\
    \email{martin.hachet@inria.fr}\\
}

\maketitle

{\small $^{*}$Co-first authorship, both authors contributed equally to this work.}

\begin{abstract}
We introduce Teegi, a Tangible ElectroEncephaloGraphy (EEG) Interface that enables novice users to get to know more about something as complex as brain signals, in an easy, engaging and informative way.
To this end, we have designed a new system based on a unique combination of spatial augmented reality, tangible interaction and real-time neurotechnologies. With Teegi, a user can visualize and analyze his or her own brain activity in real-time, on a tangible character that can be easily manipulated, and with which it is possible to interact. An explorative study has shown that interacting with Teegi seems to be easy, motivating, reliable and informative. Overall, this suggests that Teegi is a promising and relevant training and mediation tool for the general public.

\end{abstract}



\category{I.2.6 }{Learning}{Knowledge acquisition}

\begin{figure}[h!]
\begin{center}
\includegraphics[width=0.9\columnwidth]{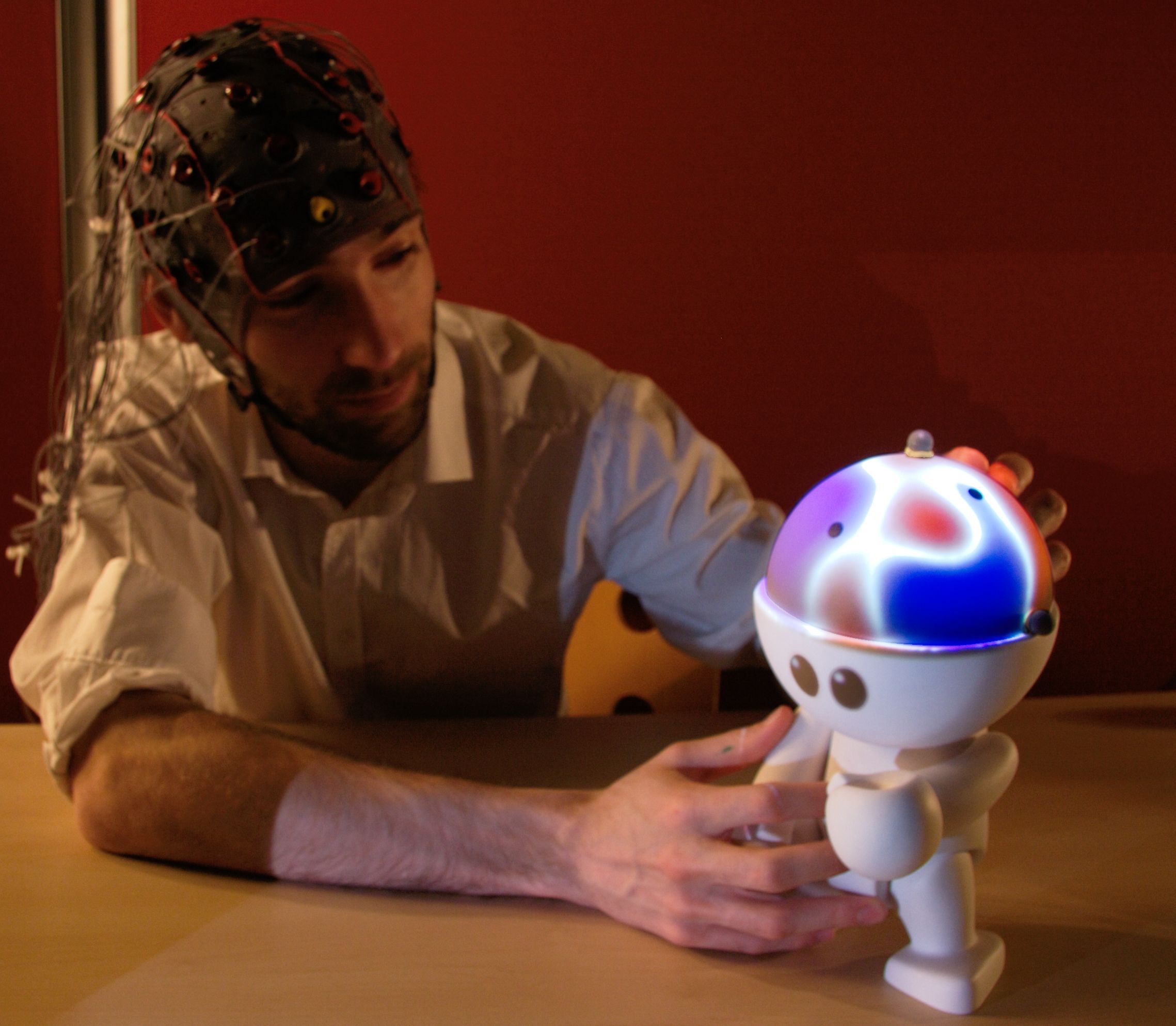}
\caption{Teegi (Tangible EEG Interface) is a friendly interactive character that users can manipulate to observe and analyze their own brain activity in real-time.}
\label{fig:trailer}
\end{center}
\end{figure}

\section{Introduction}
\label{sec:introduction}

Electroencephalography (EEG) measures the brain activity of
participants under the form of electrical currents, through use of a set of
electrodes connected to amplifiers and placed on the scalp \cite{Niedermeyer05}. This technology is widely used in
medicine for diagnostic purposes. It is also increasingly explored in the
field of Brain-Computer Interfaces (BCI), the goal of which is to enable a user
to send input commands to interactive systems without any physical motor
activities, by using brain activity alone \cite{Wolpaw12}. BCI is an
emerging research area in Human-Computer Interaction (HCI) that offers
new opportunities for interaction, beyond standard input devices~\cite{Tan10}. These emerging technologies are becoming
increasingly more popular. It feeds into fears and dreams in the general public
where many fantasies are linked to a misunderstanding of the strengths
and weaknesses of such new technologies. \emph{No, it is not possible to
read thoughts!} But what can be done exactly? Our motivation is to provide 
a tool that allows one to better learn how EEG works, and to better understand the kinds of brain
activity that can be detected in EEG signals. Beyond the knowledge of
the brain that a user can acquire, we believe that a dedicated tool may
help demystify BCI, and consequently, it may favor the development of
such a promising field.

We followed a multidisciplinary approach, combining Human-Computer
Interaction (Spatial Augmented Reality, Tangible User Interfaces),
Neurotechnologies (EEG, brain signal processing) and Psychology/Human
sciences (Human Learning and Representations, Scientific Mediation) to
design an interactive multimedia system that enables novice users to get to know more about something as complex as EEG signals and the brain, in an easy, engaging and informative way.
Our final goal is to enhance learning efficiency and knowledge acquisition
by letting users actively and individually manipulate and investigate
the concept to be learned \cite{Vosniadou01}, i.e. EEG signals.

This gave birth to Teegi (Tangible EEG Interface), a physical character
that users can manipulate in a natural way to observe and analyze their own brain
activity projected in real-time on the character's head (see Figure~\ref{fig:trailer}).
Beyond the technical description of Teegi, this paper depicts an explorative study we
conducted, which provides an experimental basis for discussions and future works. Our
major contribution for this paper is the design of the first system
to make EEG signals and brain activity easily accessible, interactive
and understandable. This work is based on theoretical foundations,
technical developments, and preliminary investigations.

\section{Neuroimaging and EEG}
\label{sec:neuroimaging-and-eeg}

EEG signals are small electrical currents (in the $\mu$V range) that can
be measured on the surface of the scalp \cite{Niedermeyer05}. They
reflect the synchronous activity of millions of neurons from the brain
cortex (i.e., the outer layer of the brain). Compared to alternative
neuroimaging techniques, such as MagnetoEncephaloGraphy or
functional Magnetic Resonance Imaging, EEG is simultaneously cheap, 
portable and provides good time resolution. 
Because of these advantages, EEG has been used for many
years in medicine, e.g., for the diagnosis of sleep disorders
or epilepsy \cite{Niedermeyer05}. More recently, with the advance of
computer processing performance, it became possible to measure and
analyze in real-time the content of these EEG signals. This paved the way for the rise  
of BCI which uses real-time analysis and decoding of EEG signals
in order to identify the mental state of the user and translate it into
a command for an application \cite{Wolpaw12}.

\begin{figure}[htbp]
\centering
\includegraphics[width=1\columnwidth]{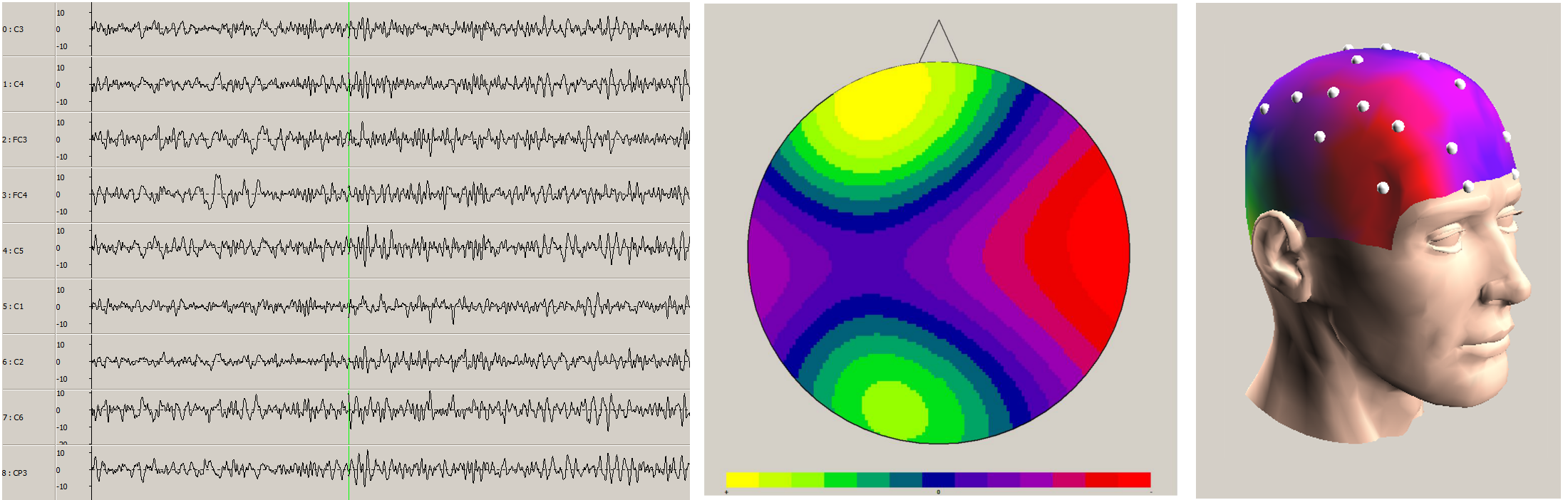}
\caption{(Left) A trace of EEG signals collected from multiple sensors.
2D (center) and 3D (right) topographic maps. (Screenshots from OpenViBE \protect\cite{Renard10}). The first two views are traditionally used by experts.}
\label{fig:openvibe}  
\end{figure}

The currently available tools used to visualize and analyze such signals are tailored for experts with a deep understanding of the
brain, EEG principles and EEG signal processing \cite{Niedermeyer05}.
Figure~\ref{fig:openvibe} (left and center) shows some typical visualizations 
of EEG signals used by experts, i.e., EEG signal traces and a 2D topographic map. 
More complex visualizations have been proposed, such as 3D topographic maps 
(Figure~\ref{fig:openvibe}, right), but they require many mouse inputs to be observed 
from all angles, which make them inconvenient to use in practice.

Although EEG visualizations are intended for experts only, 
the general public is often compelled by how the brain
works and how its activity is measured. Anyone wondering about 
brain injuries, epilepsy, sleep or learning disorders, aging,
etc. may want to seek further knowledge about how the brain works.
Currently, the public is increasingly exposed to
neurotechnologies due to the availability of consumer grade EEG
devices, such as the Emotiv EPOC or the Neurosky MindWave. Consequently, it has
become necessary to design tools and user interfaces which will allow the general
public to visualize, understand and interact with EEG signals. For
instance, Mullen \emph{et al.} proposed a software solution to process EEG
signals collected using wearable EEG devices and visualize them in 3D
\cite{Mullen13}. This software enables the user to estimate brain activity sources and connectivity but is still mainly designed for
brain signal and neuroscience experts, and not so public-friendly. 
Another recent work, more suited to lay persons, is
the ``Portable Brain Scanner'' \cite{Stopczynski14}. This system makes use of a consumer-grade EEG device (the Emotiv EPOC) and a smartphone to provide a cheap and portable solution enabling anyone to visualize the sources of their brain activity on their smartphone in 3D. Another more attractive work, which is the most closely related to
our Teegi system, is the ``Mind-Mirror'' system
\cite{mercier-ganady_mind-mirror:_2014}. This system combines Augmented
Reality (AR), 3D Visualization, and EEG to enable users to visualize
their own brain activity in real-time superimposed to their own
head, thanks to a semi-transparent mirror-based AR setup.

This short review of the existing literature about making EEG accessible
to the general public revealed that this is still a vastly unexplored
area. Moreover, these solutions do not take into account any
representation that the general public may have regarding the brain and
EEG signals -- many lay people do not even know what EEG signals are --
in order to provide suitable visualizations and interaction devices
to better understand these concepts. Some rare studies have indicated that
misconceptions about brain functions prevail in general public
\cite{Duschl12,Herculano02,Simons11}. These works stress the importance of
popular scientific communication and indicate that communication
efforts should be focused on increasing public awareness. It
is important to note that the existing works mentioned above are mostly
centered on visualization, with little or no interaction
possibilities to manipulate and understand the EEG signals in real-time
and in a friendly way. This further deters the general public from understanding brain activity \cite{Vosniadou98}. Therefore,
with the aim to enhance general public awareness, our work associates
technical innovation and user-centered design.

\section{Introducing Teegi}
\label{sec:teegi}

\subsection{Founding principles}
\label{ssec:founding-principles}

Design choices were made
according to pedagogical principles. It has long been recognized that
learner-centered education is much more effective than
transmission-based education, even in informal situations
\cite{Wellington90}. According to the constructivist paradigm, people
create unique personal meanings by reflecting on interactive learning
experiences. Therefore, people/learners should investigate and
manipulate in order to become conscious of complex phenomena, change
their misconceptions and construct scientific knowledge
\cite{Vosniadou01}. In association, meaningful models play an important
role in this type of learning processes \cite{Fleck13}. 
This motivated the design of an anthropomorphic interface that can be freely manipulated.

Our user-centered interactive media uses Spatial Augmented
Reality (SAR) and tangible interaction.
SAR, introduced by Raskar \emph{et al.}
\cite{Raskar2001a}, adds dynamic graphics to real-world
surfaces by the means of projected light. Many systems were designed
using projectors to add ``painted'' surface \cite{Raskar2001a} or to
give the illusion of virtual elements actually being there
\cite{Wilson2012,Benko2012}. A related approach is Tangible User Interface (TUI). TUI is concerned with providing tangible (i.e. physical) representations to digital information and controls \cite{Shaer2009}. One of the strengths of tangibles is their
situatedness: the interaction takes place in a real-world environment
that often hides most of the technological aspects to expose physical
interaction components only. They are particularly well suited for
mediation purposes as they tend to be more inviting compared to
mouse-screen based interfaces \cite{Horn2009}.

SAR and TUIs are often found together
\cite{Underkoffler1999,Piper2002}. They are very complementary in
that they both take place in the real world, in a common canvas. The
tangible interaction allows for a hands-on approach by offering
different input affordances (as well as physical constraints) to the
user while the SAR technology allows for a flexible and situated way to
give feedback. SAR can also be used as an affordable way to embed
dynamic graphics on a physical surface that would otherwise require
curved displays \cite{Brockmeyer2013} or rear projection \cite{Benko2008}.

There are examples of systems that use either tangible or AR principles
to interact or review physiological data. Hinckley \emph{et al.}
\cite{Hinckley1994} designed a system which used tangible props to do
neurosurgical planning. A small tangible head was used in conjunction
with a plastic plane to select the cutting planes to be visualized on a
screen. Also mentioned above, the ``Mind-Mirror''
\cite{mercier-ganady_mind-mirror:_2014} is the work closest to Teegi.
However, with Teegi, the data is not co-localized with the data source.
It provides flexibility and easier visualization as the users can change
viewpoints by tangible interactions instead of rotating their head while
keeping their eyes on the mirror. This ``out-of-body'' visualization
also enables collaboration where multiple users can explore the data.

\subsection{General description}
\label{ssec:general-description}

Teegi is a tangible interface that enables users to visualize and analyze a representation of their own brain activity recorded via an EEG system in real-time and displayed on a physical character. After some processing of the
raw signals, a dedicated visualization is projected directly on top of
the character. This character is tracked, which allows us to co-locate
the projection with the character's head, at any time. Hence, the user
can easily visualize a realistic modeling of the EEG signals in any part
of the scalp by manipulating the character, while maintaining a good
spatial topology of the observed data. Teegi was purposely given a
child-like appearance, as well as animated eyes (also projected) that
blink at the same time as the users do, in order to breathe life to the
character and enhance attractiveness. Indeed blinking can be easily detected in electrodes neighboring the eyes.

\begin{figure}[h!]
\begin{center}
\includegraphics[width=1\columnwidth]{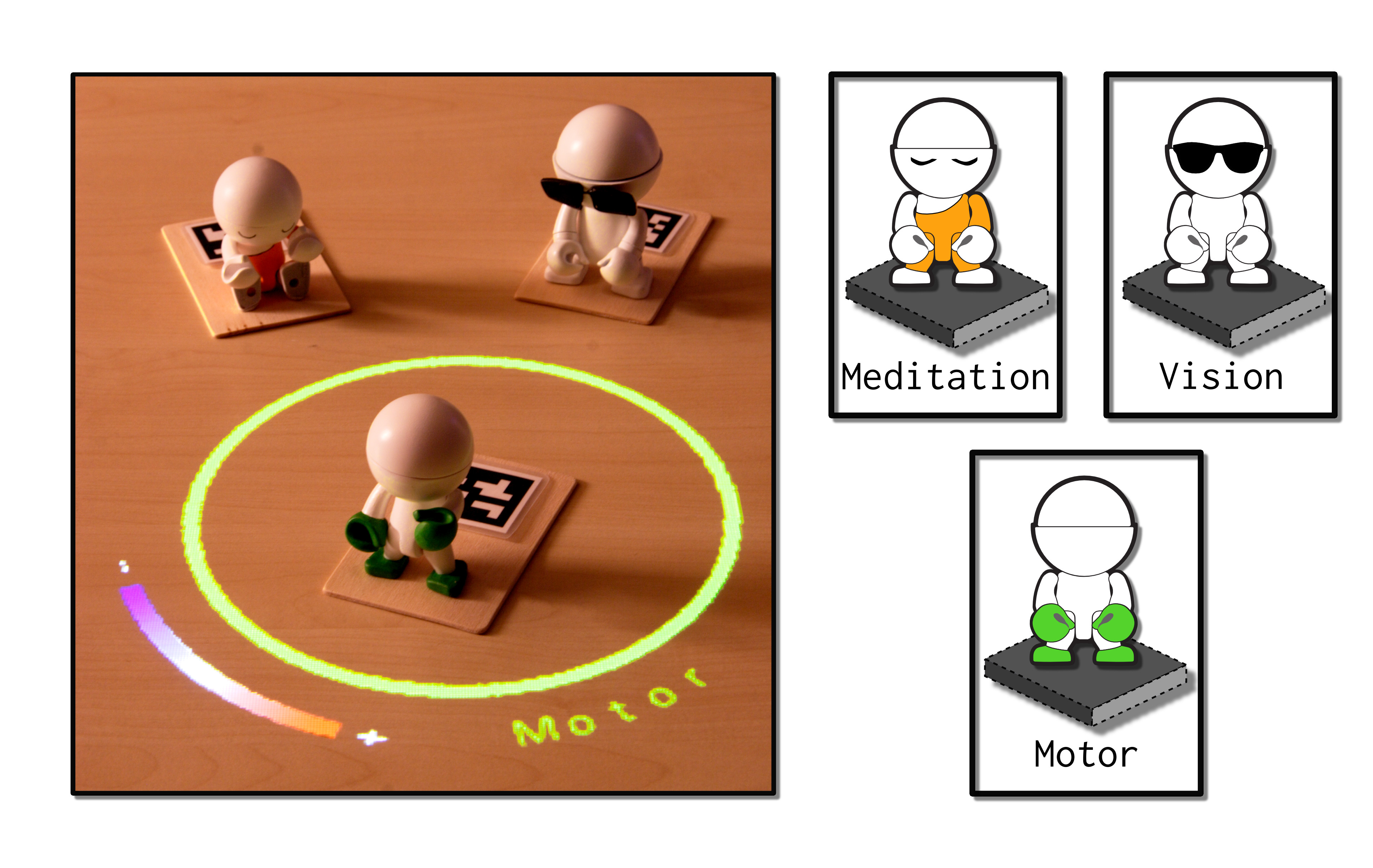}
\caption{Three mini-Teegis can be used to apply high-level EEG filters to highlight brain processes associated to \emph{Motor}, \emph{Vision} and \emph{Meditation} 
activities. To do so, the user simply needs to move the desired mini-Teegi into a specific zone projected on the table (green circle).}
\label{fig:mini-teegis}
\end{center}
\end{figure}

\begin{figure}[h!]
\begin{center}
\includegraphics[width=1\columnwidth]{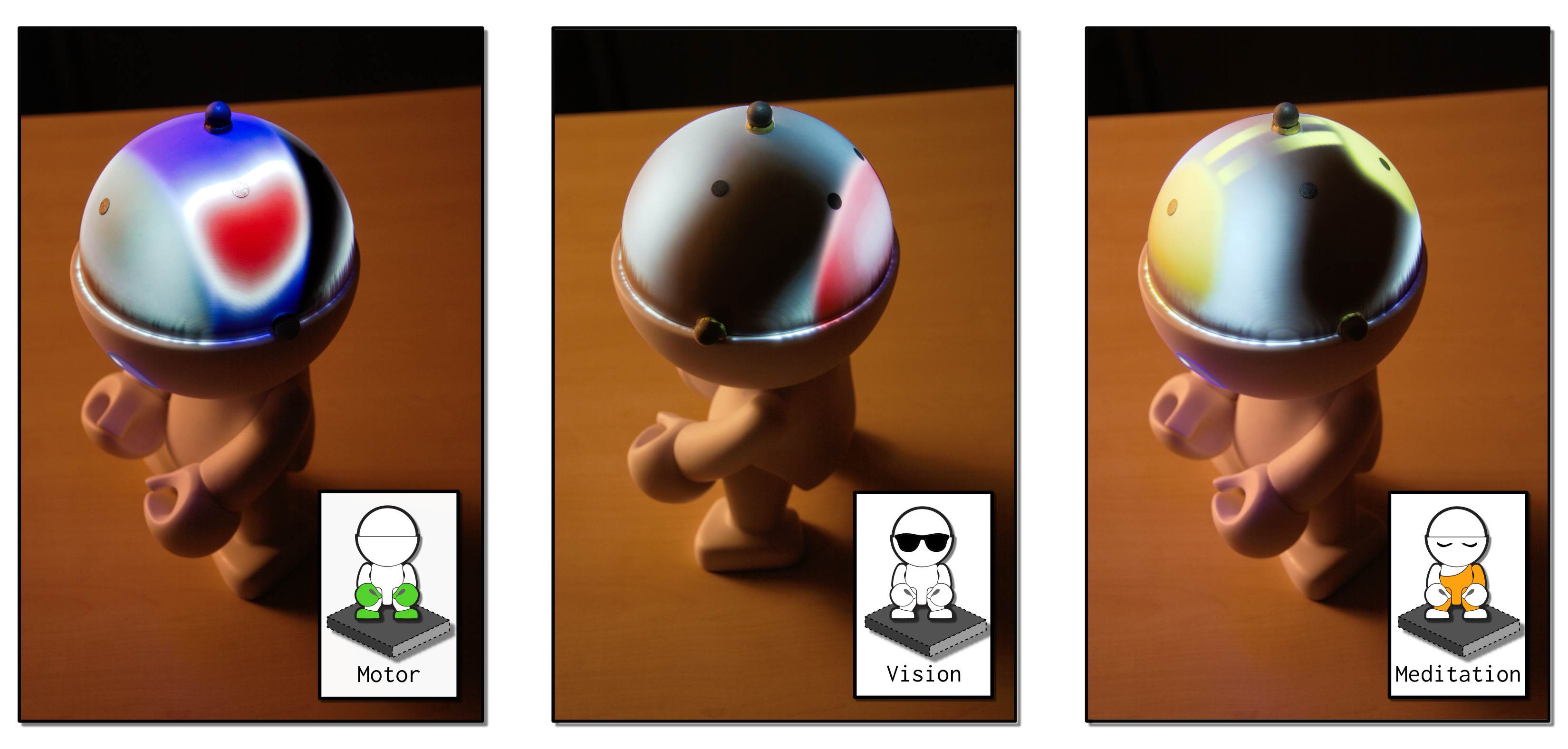}
\caption{Examples of the displayed visualizations on Teegi for each of the provided filters. 
Once a filter is active, the brain area corresponding to the selected and processed activity is highlighted in colors while the remaining EEG signals are displayed in grayscale.}
\label{fig:filters}
\end{center}
\end{figure}

Three different filters can be applied to the raw data (see the
technical section for details) enabling users to investigate influences of motor
motions, visual activities or meditation, on their brain activity in
real-time. To remain consistent with the tangible philosophy of this
project, we decided to control the filters by way of small tangible
characters (mini-Teegis) that can be moved on a ``filter area'', which
is highlighted on the table by a projected halo (see Figure~\ref{fig:mini-teegis}). For
example, if a user wants to apply a filter that will allow her to better
see what happens when moving her hand, she just needs to take the
dedicated mini-Teegi, i.e.~the one with the colored hands, and to move
it to the filter area. Then, by moving her right hand, she should see
changes in EEG amplitude on the left hemisphere of Teegi's head, as
illustrated in Figure~\ref{fig:filters}.
The manipulation of Teegi requires a motor activity. Therefore, when the motor filter is on, 
manipulating Teegi will obviously lead to observable changes in brain activity.

At the end of this paper, we present an explorative study we conducted to obtain feedback about the main features of Teegi. However, Teegi is not limited to these first features. In the next section, we describe additional interaction metaphors we have explored, and that may benefit more advanced users. These advanced features were not evaluated during the study.

\subsection{Advanced features}
\label{ssec:advanced-features}

\begin{figure}[h!]
\begin{center}
\includegraphics[width=1\columnwidth]{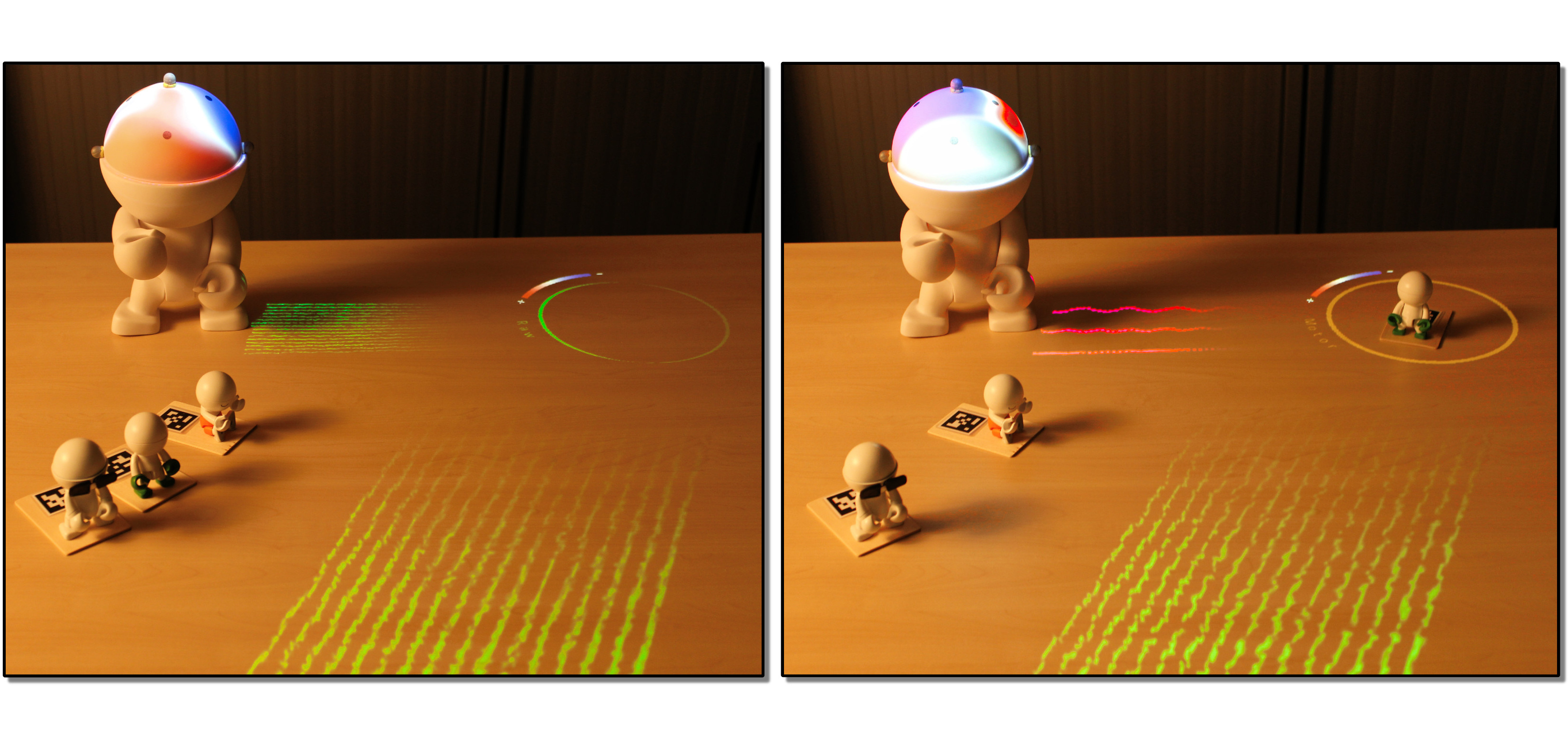}
\caption{(Left) The raw EEG readings are displayed going from the user to the filter area and then rerouted towards Teegi. (Right) When a mini-Teegi (\emph{i.e.} a filter) is active, the corresponding filtered signals are displayed between the filter area and Teegi instead.}
\label{fig:signal-table}
\end{center}
\end{figure}

Visualizing the raw signal recorded on each electrode of the EEG is not
very informative for the general public. However, this
can be instructive for students who are learning EEG signal processing and
analysis. In our approach, we can display on the table these raw data,
as shown in Figure~\ref{fig:signal-table} (left). This creates a visual link between what is
recorded with the EEG system, and the visualization that is provided on
Teegi's head. This is possible because we know the rough position of the
user, and the exact position of Teegi. When applying a filter, as
described in the previous section, the user can see the effect of his or
her action on the signal (see Figure~\ref{fig:signal-table}, right). Compared to a standard
approach where everything takes place on a screen, we believe that such
a spatial and tangible approach might ease the understanding of the filters' effect.

\begin{figure}[h!]
\begin{center}
\includegraphics[width=1\columnwidth]{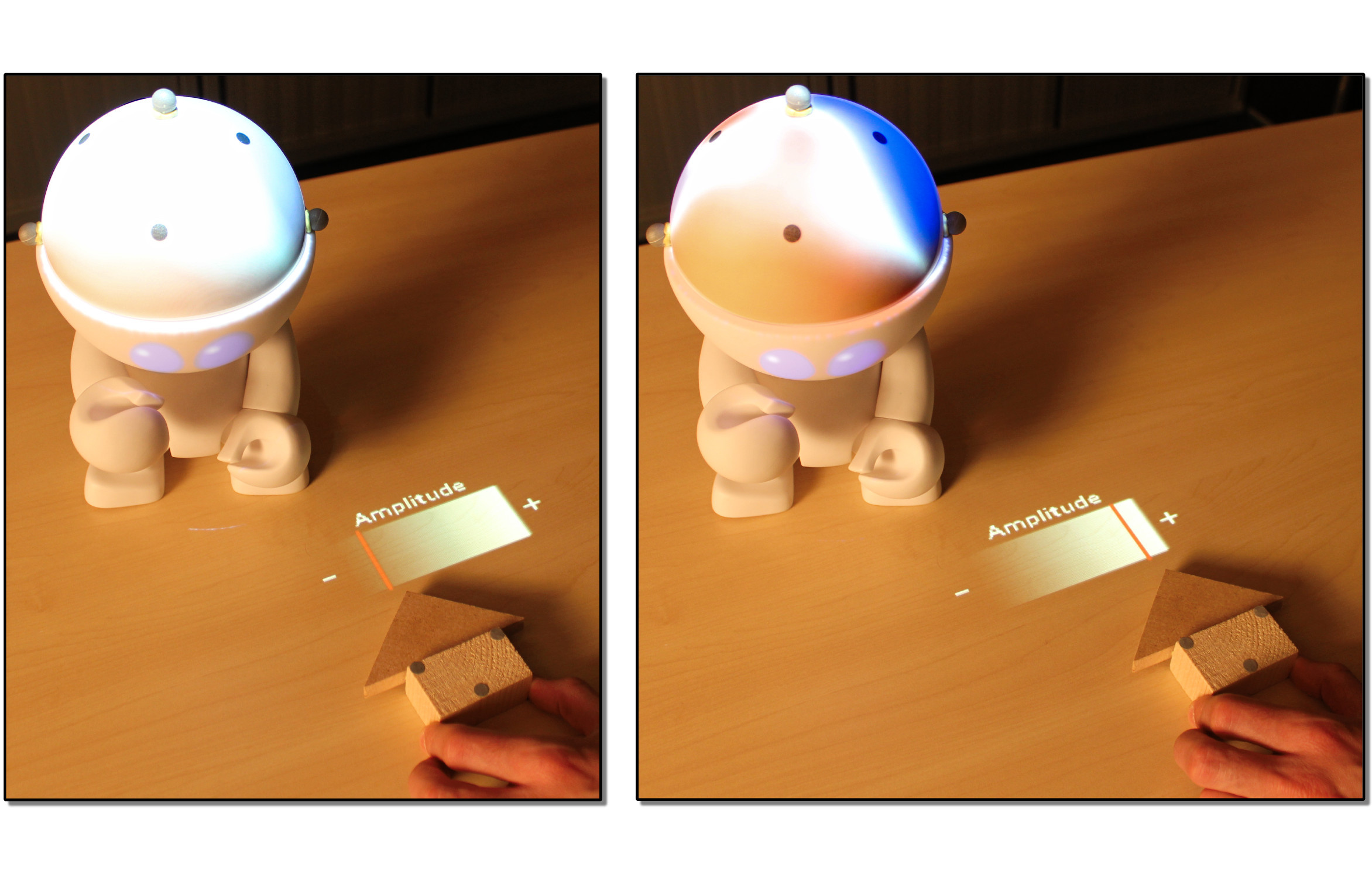}
\caption{A moving tangible cursor is controlling the amplitude of the visualization color map.}
\label{fig:tangible-threshold}
\end{center}
\end{figure}

Another dimension we explored is the use of tangible actions to control
some parameters of the EEG signal processing. As an example, we have implemented a technique where the user can control the amplitude of the visualization color map by moving a tangible object on the table (figure~\ref{fig:tangible-threshold}). This could be useful to reveal tiny fluctuations of EEG signals. With such interaction techniques, the whole analysis could be conducted without the use of a screen or a mouse, which remains consistent with the tangible philosophy of the project.

\begin{figure}[h!]
\begin{center}
\includegraphics[width=0.6\columnwidth]{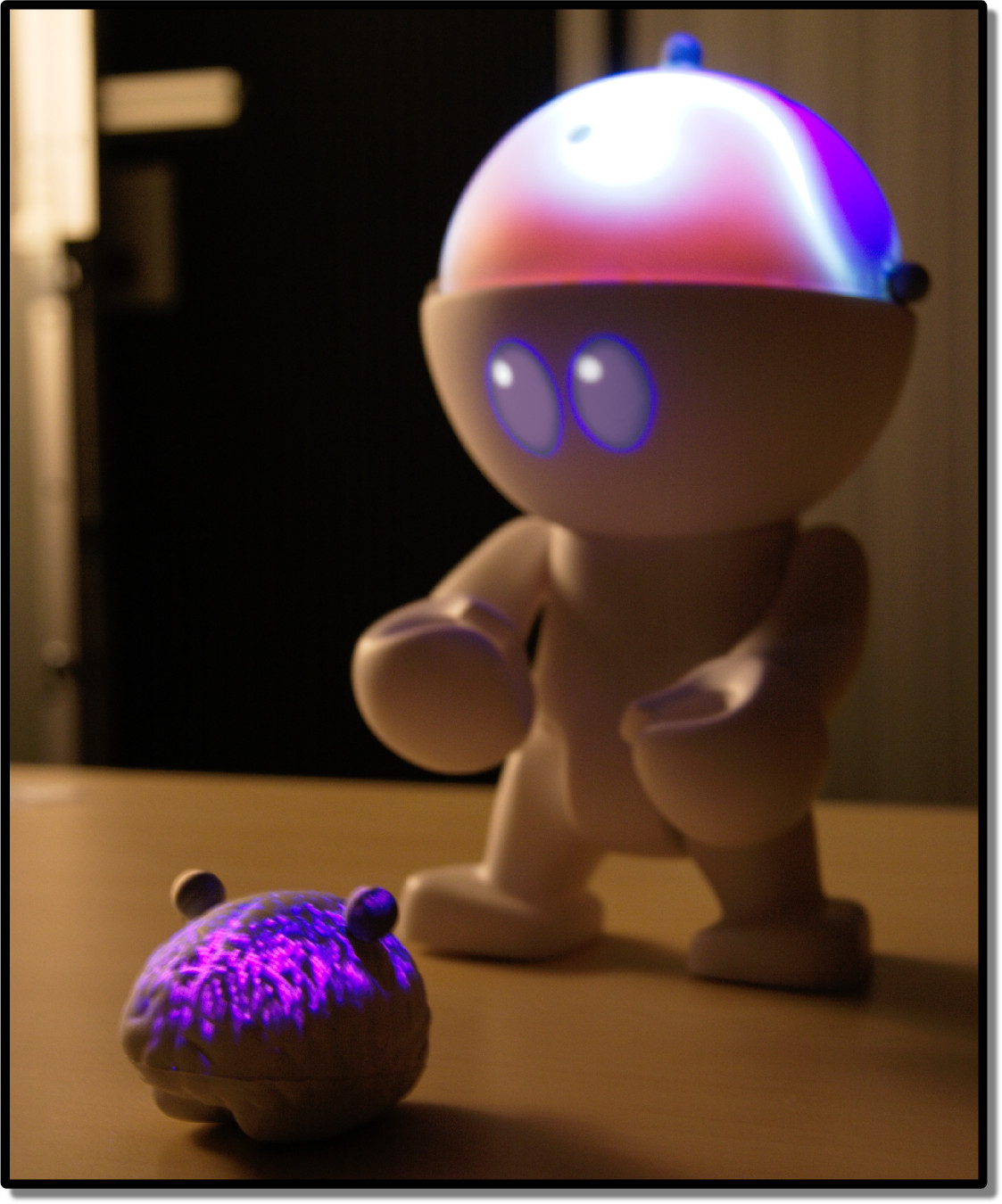}
\caption{Using an inverse model, the cortical activity and EEG measures are presented together to users.}
\label{fig:inverse-model}
\end{center}
\end{figure}

Finally, we developed a solution that highlights the relationship between EEG signals and localized 
cortical sources, that is where the signals come from \emph{inside} the brain. 
Using sLORETA inverse modeling \cite{Pascual-Marqui02} and Brainstorm to compute the 
kernel matrix \cite{Tadel2011}, we obtained a model of the cortex containing 2002 voxels 
linked to the 32 EEG electrodes we used. We can then project in real time the activity
 which arises from the outer regions of the cortex on an object representing the brain, 
 alongside with Teegi (Figure~\ref{fig:inverse-model}). Since both Teegi and the brain proxy 
 are tracked, it becomes possible to manipulate two synchronized representations of the same 
 brain activity (the source at the surface of the brain and the measures on the scalp).
  This opens way to mediation activities that are more advanced all the while keeping 
  the simplicity and ease of use brought forth by using SAR and tangible interaction.

\section{Technical description}
\label{sec:technical-description}

\subsection{EEG}
\label{ssec:eeg}

We designed different EEG signal processing pipelines that each create a specific
visualization tailored to identify specific elements in the signal. The
details of these pipelines are transparent to the user. Each pipeline corresponds to a mini-Teegi filter. In particular, we set up the
following EEG signal processing pipelines:

\begin{enumerate}
\def\labelenumi{\arabic{enumi}.}
\itemsep1pt\parskip0pt\parsep0pt
\item
  Wide-band EEG activity: EEG signals were band-pass filtered between 3
  Hz and 26 Hz, in order to filter DC drift and part of the artifacts
  (e.g., facial muscle activity \cite{Fatourechi07}) that may pollute
  them. Their power is then computed before being displayed. This corresponds to unspecific brain signals, hence they were labeled as "raw" signals.
\item
  Sensorimotor activity: EEG signals were first band-pass filtered in the
  $\beta$ band (16-24Hz), a brain rhythm highly involved in sensorimotor
  tasks \cite{Pfurtscheller99b}. Then, they were spatially filtered, i.e.,
  the signals from several neighboring EEG sensors were combined in order
  to enhance the signal of interest. In particular, we used and displayed
  Laplacian spatial filters around electrodes C3, C4 and Cz. This enabled the users to visualize EEG activity
  changes due to movements of the left hand, right hand and feet.
  Indeed, it is known that the power of EEG signals in the $\beta$
  rhythm decreases in electrodes C3/Cz/C4 during right hand/feet/left
  hand movements respectively, and increases just after the end of this
  movement \cite{Pfurtscheller99b}.
\item
  Visual activity: EEG signals were band-pass filtered in the $\alpha$
  band (8-12 Hz), then only electrodes P3, Pz, P4, PO3, POz, PO4, O1, Oz
  and O2 (located on the back of the head, above the neck) were selected
  and displayed. These electrodes are indeed located over the visual
  cortex of the brain, i.e., the brain area in charge of visual
  information processing. The amplitude of the $\alpha$ rhythm is actually 
  known to increase while the user is closing his/her eyes and is thus
  not processing any visual information \cite{Niedermeyer05}. To ensure that the user could perceive this increase after he/she reopened his/her eyes, the visualization was delayed by 0.5s.
\item
  Meditation: on a more exploratory note, we used the synchronization between the signals from the anterior and posterior cortex (AFz/Pz), which
  was measured in a 7-28 Hz band with instantaneous phase locking value \cite{lachaux_studying_2000}.
  There are different outcomes (increase/decrease in synchronization)
  depending on meditation type. Mindfulness and body focus practices decrease the synchronization while
  transcendental practice increases it \cite{lehmann_reduced_2012}.
\end{enumerate}

EEG signals were acquired with a 32-channels EEG device (made of two g.tec g.USBAmp EEG amplifiers).
This professional-grade system ensured that our prototype had a good signal-to-noise ratio and accurate electrode location, avoiding unneeded uncertainties.
Signals were processed in real-time using OpenViBE \cite{Renard10}. For pipelines 1 to 3, 
the displayed colors correspond to signal power strength; for pipeline 4 they correspond to the
degree of synchronization.

\subsection{Spatial Augmented Reality}
\label{ssec:spatial-augmented-reality}

\begin{figure}[h!]
\begin{center}
\includegraphics[width=1\columnwidth]{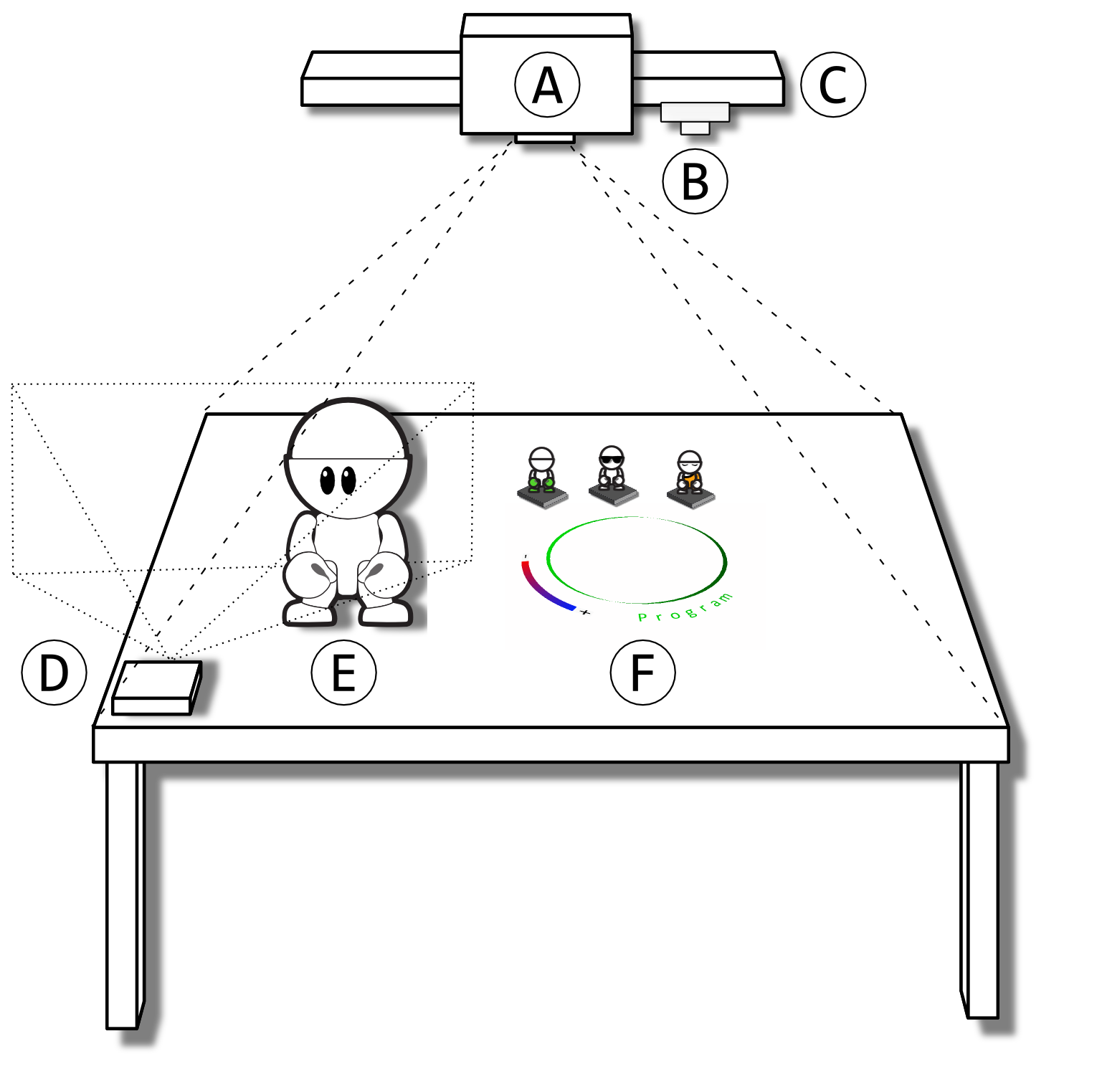}
\caption{Diagram of the installation. (A) ProjectionDesign F20SX projector (B) Sony PSEye web camera (C) OptiTrack V120:Trio (D) Vivitek Qumi Q2 projector (E) Teegi (F) Program selection zone and mini-Teegis.}
\label{fig:setup-projection}
\end{center}
\end{figure}

In order to create an augmented character, we have designed a tabletop
augmentation setup (see Figure~\ref{fig:setup-projection}). Teegi itself is a 25cm high Trexi DIY toy. The
mini-Teegis are also 10cm high Trexis. The main program handling the
whole installation was created with
vvvv \cite{vvvv}. The primary projected content (Teegi augmentation and
GUI display) is handled with a single wide lens projector
ProjectionDesign F20SX of resolution 1024x768 located over the table in
a top-down orientation. The tracking of Teegi is achieved with an
OptiTrack V120:Trio. It runs at 120 FPS with an overall latency of 8.3ms
and a precision of 0.8mm. The OptiTrack is located in the same
configuration as the main projector and both devices are calibrated
together manually. The tracking data is sent to vvvv
using OptiTrack's NatNet protocol. Teegi's eyes are projected using
a second projector (Vivitek Qumi Q2) that is located on the side of the
table.

The filter selection is done using a Sony PSEye web camera pointed at
the position of the program selection GUI. Each mini-teegi representing
a filter has a fiducial marker attached to it. The
library ARToolkitPlus \cite{Wagner2007} is used to detect which marker is
currently selected.

The EEG signals are processed by the OpenViBE software \cite{Renard10}
that also generates a grayscale texture of the scalp signals. This texture is then
exported to a local shared samba folder which is then fetched and remapped to an appropriate color scale in vvvv
before being mapped to Teegi's head. In addition, the raw EEG signals
are sent to vvvv over VRPN for display purposes (see Figure~\ref{fig:signal-table}).

\section{Explorative study and discussions}
\label{sec:evaluation}

\subsection{Protocol}
\label{ssec:protocol}

We conducted an explorative study where participants had to manipulate Teegi
following a given scenario. The objectives of this study were to i)
evaluate the general usability of the interface and ii) obtain initial
feedback about the relevance of the approach to help users understand EEG signals and the brain. Ten participants (6 males, 4 females,
mean age 28.6 (SD=9.7)) took part in this study. Pre-tests
confirmed they were rather naive on the subject.
They manipulated the version of Teegi described in the General Description section (no advanced features). The general procedure was as follows:

\begin{enumerate}
\def\labelenumi{\arabic{enumi}.}
\itemsep1pt\parskip0pt\parsep0pt
\item
  Pre-tests: The participant answered a first questionnaire assessing his or
  her representation of the brain. The participant then filled in different
  forms to measure his or her previous knowledge; one form per studied
  brain process (motor, vision and meditation).
\item
  Setting-up: The experimenter positioned the EEG cap on the participant's
  head. In parallel, the participant, guided by the experimenter, was made aware of the four
  didactic ``cards'' explaining the different filters \emph{i.e.},
  \emph{Motor}, \emph{Vision}, \emph{Meditation} and \emph{Raw}. Each card was comprised of an image of the mini-Teegi associated with the filter along with basic instructions to follow (e.g. the \emph{Motor} card indicated to the participants to move their hands or feet while staying relaxed). There were also two cards describing the two types of
  visualization participants could face, \emph{signal strength} and
  \emph{synchronization}. Once the participant was equipped,  a quick
  calibration phase occurred. While Teegi was still inactive, participants were asked to close their eyes for a few seconds, and to move their hands and feet in order to identify the baseline activity for visualization.
\item
  Personal Investigation: The participant was asked to freely manipulate 
  Teegi as well as the filters to be able to answer the following
  questions:

  \begin{itemize}
  \itemsep1pt\parskip0pt\parsep0pt
  \item
    What happens when you move your hands or feet?
  \item
    What happens when you close your eyes?
  \item
    What happens when you meditate?
  \end{itemize}

  During the whole study the participant sat comfortably in a chair. To avoid the occurrence of muscle artifacts that may pollute the signals, the user was instructed to stay relaxed and to refrain from making strong head movements.
  
\item
  Post-tests: The participant answered the questions above on dedicated
  forms, the same that were given at the beginning of step 1. Finally, he or
  she filled in a user survey questionnaire based on a 7-point Likert scale.
\end{enumerate}

The whole session lasted approximately 1.5 hours per participant,
with 15 to 20 minutes of hands-on time with Teegi. Each
session was video-recorded. Video segments were separately visualized and labeled with the corresponding behavior (i.e. tangible and visual interactions,
emotional expressions, and investigation strategies) using The Observer
XT{\textregistered} 11.5 (Noldus, Info Tech, Wageninen, The Netherlands). After the session, the experimenter had an informal talk with the participant. He corrected the answers, making sure the participant was not leaving with false knowledge, and explained in more detail some aspects of the system (e.g. relationship between visual filter and attentional states, the various effects of meditation, ...). This phase lasted from 30 min to 1 hour depending on the participant's curiosity.

\subsection{Results and discussions}
\label{ssec:results-and-discussions}

\begin{figure}%
\begin{center}
\includegraphics[width=\columnwidth]{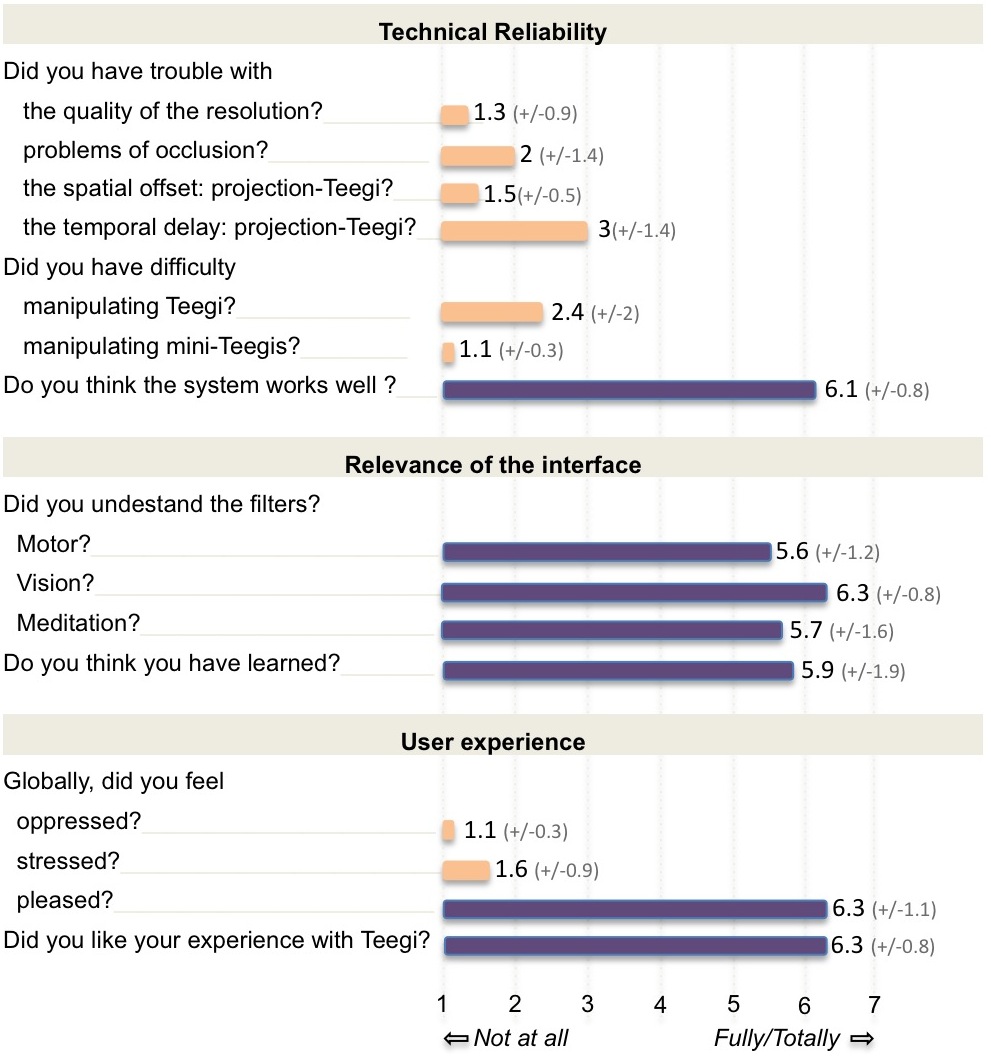}
\caption{Results of the questionnaire (selected questions). Note that purple (resp. orange) bars indicate questions measuring Teegi's qualities (resp. limitations).}%
\label{fig:likert-data}%
\end{center}
\end{figure}

To better understand the inherent strengths of Teegi
towards learning, we assessed three main aspects of Teegi: its technical
reliability, its relevance to ease understanding for non-experts, and the User eXperience (UX) it provides. This evaluation is based on 
1) the results of the questionnaire that are summarized in Figure \ref{fig:likert-data}, 
2) the analysis of the video recordings, and 
3) the analysis of the forms the participants filled in to assess their pre and post-knowledge of the brain and EEG.

\subsubsection{Technical reliability}
\label{sssec:technical-reliability}

Participants unanimously reported that the whole system worked properly. 
The quality of the SAR display is valued by the participants. In particular, they reported that the resolution was appropriate, 
and they did not report problems of offset between the display and the physical character.
Participants declared that they were not disturbed by occlusion problems. 
The mild temporal delay between their action and their consequences seems not to be an issue.

Manipulations of Teegi were numerous and frequent. 
Teegi was touched or moved on average 25\% of the session's duration, twice per minute. 
These manipulations consisted mostly of rotations, and to
a lesser extent of lifting Teegi to enhance visual perception.
Two participants reported difficulties in grasping Teegi while the remaining 8 were comfortable 
with the form of the character. Video analyses did not
show difficulties for the manipulation of Teegi. 
Similarly, applying filters by manipulating the mini-Teegis seemed easy for the participants.

\subsubsection{Relevance of the interface to ease understanding}
\label{sssec:relevance-of-teegi}
  
The participants reported that they understood the visualization associated with
the filters. Video analyses indicated that they systematically used
all filters several times (3 times per session on average) for a similar duration (Raw filter : 30.4\% (SD 13.3) of
session duration; Motor filter: 26.0\% (8.3); visual filter: 16.9\%
(5.5); meditation filter: 26.6\% (8.7)). Interestingly, the visual activity filter seemed
slightly easier to understand than the other filters. Moreover, video
analyses indicated that the participants did not have difficulty observing the signals on Teegi's head, 
as soon as they found the right location to observe. Overall, participants reported that they were able to use Teegi without any difficulties. 

\begin{figure}[h!]
\begin{center}
\includegraphics[width=0.9\columnwidth]{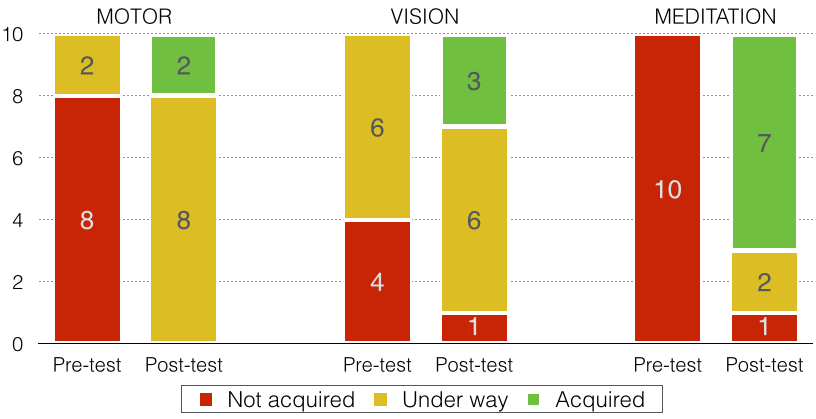}
\caption{Marks obtained by the participants during the pre- and post-test assessments. See text for details.}
\label{fig:assessment-teegi}
\end{center}
\end{figure}

All participants completed the required tasks. They used instruction cards 5 times
per session on average.
They reported that they could focus on the tasks rather than on the mechanisms used to
achieve them. This suggests that Teegi is a rather transparent interface. 
Regarding learning of brain processes and EEG, participants
reported that they believed they had learned  while doing the study. 
This was confirmed by the results of the pre- and post-test assessments (see Figure \ref{fig:assessment-teegi}). These assessments focused on the recognition and the understanding of brain activation during Motor activities, Visual activities and Meditation. Understanding was marked as acquired if 1) the activated areas were correctly localized and 2) the explanations of the brain process were correct. It was marked as under way if only 1) or 2) was satisfied but not both; and as not acquired if neither 1) nor 2) were satisfied. The marks obtained by the participants improved after using Teegi. Overall, this suggests that Teegi offers many interesting features to ease learning and mediation.

All our results indicate that Teegi clearly promotes real-time tangible interactions,
which contributes to enhancing awareness. Constructivism and
inquiry-based science education principles indicate that, to
become conscious of complex phenomena and construct scientific knowledge,
people/learners have to experiment by interacting with and physically
manipulating the content \cite{Vosniadou98}. This is particularly true for 
brain activity that is difficult to understand because it cannot be sensed \cite{Damasio1994}, contrary 
to other activities (e.g. respiratory) that are perceived through sensory-motor mechanisms.
Hence, brain activities need to be conceptualized, and the success of learning processes strongly depends on the interface. Teegi, which has been largely promoted by the participants, seems to fulfill this function.

\subsubsection{User eXperience}
\label{sssec:user-experience}

The general experience with Teegi was rated as pleasant, attractive and stimulating, 
and participants did not feel stressed or oppressed. Overall, participants reported that they liked 
interacting with Teegi.  The emotion expression analyses 
confirmed those statements. They showed
that on average participants expressed curiosity and questioning about Teegi
feedback during almost 20\% (20.1\% SD=9.1) of the manipulation
duration. Other emotion expressions observed for all participants were joy
and pleasure (e.g.~smile, laugh, joyful verbal expression\ldots{}). They
occurred during almost 10\% (9.8\% SD= 6.7) of the interaction duration
with Teegi. Surprise emotions were observed but less frequently.
Interestingly, boredom, weariness expressions occurred rarely (only for 2
users) and only at the end of the manipulation time. We did not
observe any occurrence of exasperation or irritation. These results
suggest a high level of acceptance for Teegi. This is a fundamental requirement for 
a tool aiming at improving access to knowledge.

Behavior observations indicated that the majority of participants spoke with Teegi and used morphological zones specific to human interactions while manipulating it. For example, they held its hands and held it up by the waist as one would do with a child. Some users spoke in the first person when they observed changes on the character's scalp for example ``so, when I move my hands, I light up on the sides''; many said aloud that Teegi was their own image, for example ``so, Teegi is me!''. 
This identification suggests that an activation of associations between the perceived character's personality and self-perception may have occurred \cite{Paiva05}. It is known that identification can be associated with increasing loss of self-awareness, and its temporary replacement with elements of the perceived character's personality \cite{Cohen01}. Therefore a human shaped, child-like character, made lifelike by animated projected eyes, could enhance both empathy and implicit self-perception of one's own brain activity, as provided by our interactive media. The anthropomorphic appearance of Teegi could explain the motivation and positive UX reported by the users. All these hypotheses would be the aim of a more extensive UX study.

Regarding visual attention, the participants were apparently paying
attention to Teegi most of the time (83.3\%, SD 7.6).
This supports the fact that Teegi mobilized
user attention. It also indicates a cognitive user engagement. Personal
investigations were permanent (only 1.9\% of inactivity was measured during the session
duration; SD=1.7). Behavior analyses indicate that participants made
predictions, hypotheses and tested them by conducting experiments.
Numerous trial and error strategies were frequently used. This clearly indicates 
 personal active control of the task and inquiry
processes. Overall, Teegi stimulates investigations and encourages persistence in task
completion.

\section{Conclusions}
\label{sec:conclusions}

In this paper, we presented Teegi, a tangible interface that makes EEG understandable to non-expert users. 
Our main contribution is the interface itself, which is built from both theoretical foundations, 
notably from human learning and scientific mediation and technical developments, including spatial 
augmented reality, tangible interaction and real-time neurotechnologies. We also demonstrated
 that this interface was well accepted by a first pool of users. In the future, we plan to make a more in-depth investigation into how well users are able to learn about EEG and brain activity with Teegi. To this end, we will conduct dedicated experiments with students and/or visitors in scientific museums. We would also like to precisely evaluate how Teegi benefits learning  compared to standard approaches. For more advanced users, ad-hoc tangible filter creation could prove to be of great interest, adding flexibility to the overall system.
  Finally, it is known that BCI requires the user to learn to control his/her 
  own brain activity to input computer commands \cite{Wolpaw12}, which is a long and tedious task. We expect Teegi
   to be a motivating and informative way to support this training in the future.

%
%
%
%
%
\balance


\bibliographystyle{acm-sigchi}
\bibliography{biblio}
\end{document}